\documentclass[conference,a4paper,10pt]{IEEEtran}
\usepackage[left=1.53cm,right=1.53cm,top=1.4cm,bottom=1.4cm]{geometry}
\IEEEoverridecommandlockouts

\usepackage[utf8]{inputenc} 
\usepackage[T1]{fontenc}
\usepackage{url}
\usepackage{ifthen}
\usepackage[noadjust]{cite}
\usepackage[cmex10]{amsmath} 
\usepackage{amssymb}
\usepackage{graphicx}
\usepackage{epstopdf}
\usepackage{epsfig}
\usepackage{multirow}
\usepackage{amsthm}
\usepackage{comment}
\usepackage{caption}
\usepackage{enumitem}
\usepackage{amsmath}
\usepackage{color}
\usepackage{bbm}
\usepackage{algorithm}
\usepackage{algpseudocode}
\usepackage{csquotes}
\usepackage{subfiles}
\usepackage{caption, multirow, makecell}
\usepackage{array}
\usepackage{caption}
\usepackage{cuted}
\usepackage{subcaption}
\usepackage{mathtools}

\newtheorem{defn}{Definition}

\newtheorem{rem}{Remark}

\def\BibTeX{{\rm B\kern-.05em{\sc i\kern-.025em b}\kern-.08em
    T\kern-.1667em\lower.7ex\hbox{E}\kern-.125emX}}

\begin{document}

\title{Shared Cache Coded Caching Schemes with known User-to-Cache Association Profile using Placement Delivery Arrays \\

}

\author{\IEEEauthorblockN{Elizabath Peter, K. K. Krishnan Namboodiri and B. Sundar Rajan}
	\IEEEauthorblockA{Department of Electrical Communication Engineering, IISc
		Bangalore, India \\
		E-mail: \{elizabathp,krishnank,bsrajan\}@iisc.ac.in}
}

\maketitle

\begin{abstract}
	This work considers the coded caching problem with shared caches, where users share the caches, and each user gets access only to one cache. The user-to-cache association is assumed to be known at the server during the placement phase. We focus on the schemes derived using placement delivery arrays (PDAs). The PDAs were originally designed to address the sub-packetization bottleneck of coded caching in a dedicated cache setup.  
	We observe that in the setup of this paper permuting the columns of the PDA results in schemes with different performance for the same problem, but the sub-packetization level remains the same. This is contrary to what was observed for dedicated cache networks. We propose a procedure to identify the ordering of columns that gives the best performance possible for the PDA employed for the given problem. 
	 Further, some specific classes of PDAs are chosen and the performance gain achieved by reordering the columns of the PDA is illustrated.

\end{abstract}


\section{Introduction}
Caching is an effective strategy to reduce the traffic congestion experienced during peak hours in content delivery networks. The memories distributed across the network are utilized to prefetch contents during off-peak times and this is called as placement phase. The cached contents are then used to serve the demands of the users during peak times, thereby reducing the congestion in the delivery phase. In the seminal work, \cite{MaN} by Maddah-Ali and Niesen, it is shown that apart from the achievable local caching gain, coded transmissions offer an additional gain called global caching gain, which is proportional to the total cache size in the network. The network model considered in \cite{MaN} is that of a dedicated cache network where there is a server with $N$ equal-length files connected to $K$ users through an error-free shared link. Each user possesses a cache of size equal to $M$ files. The performance measure is the delivery load which is defined as the normalized size of the transmission made by the server in the delivery phase. The coded caching approach has then been extended to a variety of settings that include decentralized caching \cite{MaN2}, shared cache networks \cite{PUE, IZY, DuTh, PaE}, schemes with less sub-packetization levels \cite{YCT,TaR} and many more. 

To achieve the maximum global caching gain in dedicated cache network, it is shown that each file needs to be split into at least $\binom{K}{\frac{KM}{N}}$ parts. The number of parts or packets that constitute a file is defined as the sub-packetization level in the coded caching literature. The sub-packetization level required in the Maddah-Ali Niesen scheme grows exponentially as the network scales. This makes the practical implementation of the scheme infeasible. Later, Yan \textit{et al.} introduced the combinatorial structures called Placement Delivery Arrays (PDAs) \cite{YCT} which resulted in schemes with low sub-packetization levels and also characterized the placement and delivery phases in a single array. 

The sub-packetization level requirement of the optimal coded caching scheme \cite{PUE} for shared cache networks, where several users share a cache instead of having a dedicated one, is also exponential with respect to the number of caches, $\Lambda$. The study of shared cache networks is important as it succinctly captures more practical settings such as a transmitter communicating to a set of users with the help of cache-aided intermediate nodes, where all the users served by a particular node (that is, users present within the coverage of an intermediate node) have access to its cache contents. The sub-packetization level required in \cite{PUE} is $\binom{\Lambda}{\frac{\Lambda M}{N}}$. Even though the number of caches is less than or equal to the number of users, the value of $\binom{\Lambda}{\frac{\Lambda M}{N}}$ is significantly large for moderate values of $\Lambda$ itself. This problem was addressed in \cite{PeR} where the PDAs were leveraged to obtain schemes for shared cache systems with reduced sub-packetization levels. The shared cache schemes given in \cite{PUE} and \cite{PeR} follow a placement policy which is independent of the number of users accessing each helper cache. The number of users accessing each helper cache is given by the user-to-cache association profile. In this work, we consider the shared cache schemes obtained from PDAs and design its placement policy according to the user-to-cache association profile, thereby achieving a better performance compared to the scheme in \cite{PeR}.

 \subsection{Contributions}
 The PDA derived schemes in \cite{PeR} brought down the sub-packetization level from $\binom{\Lambda}{\frac{\Lambda M}{N}}$ to a lower value by paying in the delivery load. In a dedicated cache setting, the permutation of columns of the PDA affects the content placement, but the number of transmissions needed to satisfy the users' demands remains the same. Whereas in a shared cache network, rearranging the columns of the PDA that we use affects the cache placement and the number of transmissions needed. We refer to the PDAs that differ by column permutations as equivalent PDAs. In this work, our focus is on finding the equivalent PDA that results in the least delivery load achievable with the PDA employed in the given problem. Our contributions are summarized below.
 \begin{itemize}
 	\item For a shared caching scheme derived from PDAs, we propose a general procedure to identify the best possible ordering of the columns of the PDA by taking into account the user-to-cache association profile (Section~\ref{subsec:proc}).
 	\item In particular, we choose a class of PDAs obtained using Construction B in \cite{YCT} and show how to find the PDA from the set of equivalent PDAs that results in the best performance for the given shared caching problem (Section~\ref{subsec:proc_constB}).
   
 \end{itemize}
 
The reordering of columns of the PDA helps to reduce the delivery load without increasing the sub-packetization level.

 \begin{figure}[t!]
	\begin{center}
		\captionsetup{justification=centering}
		\includegraphics[width=0.85\columnwidth]{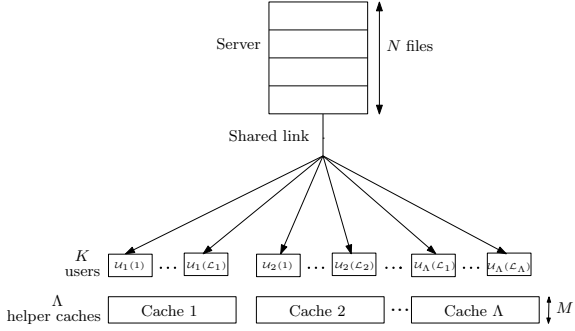}
		\caption{Problem setting for a shared cache network.}
		\label{fig:setting}
	\end{center}
\end{figure}

\subsection{Notations}
For any integer $n$, $[n]$ denotes the set $\{1,2,\ldots,n\}$. For two positive integers $m$ and $n$, $[m,n]$ denotes the set $\{m,m+1,\ldots,n\}$ and $[m,n)$ denotes the set $\{m,m+1,\ldots,n-1\}$. For a set $\mathcal{S}$, $|\mathcal{S}|$ denotes its cardinality. Bold uppercase and lowercase letters are used to denote matrices and vectors, respectively. The columns of an $m \times n$ matrix $\mathbf{A}$ is denoted by $\mathbf{a}_1, \mathbf{a}_2, \ldots, \mathbf{a}_n$. The symbol $\mathbb{N}$ denotes the set of natural numbers. The finite field with $q$ elements is denoted by $\mathbb{F}_q$. For two positive integers $m$ and $n$, $(m+n)_q$ denotes that the sum is performed under modulo $q$. 

\section{Problem Setup and Background}
In this section, we first describe the problem setup followed by a brief review on PDAs \cite{YCT} and the shared cache scheme in \cite{PeR}.

\subsection{Problem Setup}
Consider a shared cache network as illustrated in Fig.~\ref{fig:setting}. There is a server with access to a library of $N$ equal-length files $\{W^1, W^2,\ldots,W^N\}$, connected to $K$ users through an error-free broadcast link. There are $\Lambda \leq K$ helper caches, each of normalized size ${M}/{N}$, and each user gets access to exactly one helper cache. Each cache can serve an arbitrary number of users. The number of users connected to each cache is known to the server at the placement phase itself. The number of users connected to each cache, $\lambda \in [\Lambda]$ is denoted as $\mathcal{L}_{\lambda}$ and, the overall user-to-cache association profile is denoted by $\mathcal{L}=(\mathcal{L}_1, \mathcal{L}_2,\ldots,\mathcal{L}_{\Lambda})$. We assume that $\mathcal{L}$ is arranged in the non-increasing order; if not, relabel the caches accordingly. The system operates in two phases:
\begin{enumerate}[label=\alph*)]
	\item \textit{Placement phase}: In this phase, the server fills the helper caches uniformly with contents from the library of files in an uncoded form, satisfying the memory constraint. The content placement is independent of the subsequent demands of the users.
	\item \textit{Delivery phase}: After each user $k \in [K]$ gets access to one of the helper caches, the users request one of the $N$ files from the server. Let the request or demand vector be denoted as $\mathbf{d}=(d_1,d_2,\ldots,d_K)$. On receiving the demand vector $\mathbf{d}$, the server sends a message $X$ to satisfy the demands of the users. Each user is able to retrieve its demanded file using the received message and its accessible cache contents.
\end{enumerate}

Let $R(\mathcal{L},\mathbf{d})$ denote the normalized size of $X$. The worst-case delivery load required for the association profile $\mathcal{L}$ is given by $\underset{\mathbf{d} \in [N]^{K}}{\max}R(\mathcal{L},\mathbf{d})$, and is denoted by $R(\mathcal{L})$. Our objective is to design the placement and delivery policies accordingly such that the worst-case delivery load $R(\mathcal{L})$ is minimum.

\subsection{Overview on PDAs, Generalized PDA and the scheme in \cite{PeR}}
In this subsection, we first discuss PDAs followed by Generalized PDAs and then describe how Generalized PDAs represent a coded caching scheme for shared cache networks. 

\subsubsection{Placement Delivery Array (PDA)} 
\begin{defn}
	(\cite{YCT}) For positive integers $K, F, Z$ and $S$, an $F \times K$ array $\mathbf{P}=(p_{j,k})$, $j \in [F]$ and $k \in [K]$, composed of a specific symbol $\star$ and $S$ positive integers $1,2,\ldots, S$, is called a $(K,F,Z,S)$ placement delivery array (PDA) if it satisfies the following three conditions: \\
	\textit{C1}. The symbol $\star$ appears $Z$ times in each column.\\
	\textit{C2}. Each integer occurs at least once in the array.\\
	\textit{C3}. For any two distinct entries $p_{j_1,k_1}$ and $p_{j_2,k_2}$, $p_{j_1,k_1}=p_{j_2,k_2}=s$ is an integer only if
	\begin{enumerate}[label=\alph*)]
		
		\item $j_1 \neq j_2$, $k_1 \neq k_2$, i.e., they lie in distinct rows and distinct columns, and
		\item $p_{j_1,k_2}=p_{j_2,k_1}=\star$, i.e., the corresponding $2\times2$ sub-array formed by rows $j_1, j_2$ and columns $k_1,k_2$ must be of the following form:
		\begin{center}
			$\begin{pmatrix}
			s & \star\\
			\star & s
			\end{pmatrix}$
			\hspace{0.3cm}or\hspace{0.3cm}
			$\begin{pmatrix}
			\star & s \\
			s & \star
			\end{pmatrix}$ 
			
		\end{center}
		
	\end{enumerate}
\end{defn}

The PDA is said to be a regular PDA if all the integers occur an equal number of times in the array.

Every $(K,F,Z,S)$ PDA represents a coded caching scheme for a dedicated cache network with $K$ users and $\frac{M}{N}=\frac{Z}{F}$. The sub-packetization level required is $F$ ($W^n=\{W^n_1,W^n_2,\ldots,W^n_F\}, \forall n \in [N]$), and the worst-case delivery load achieved is $\frac{S}{F}$. In a $(K,F,Z,S)$ PDA $\mathbf{P}$, the rows represent subfiles and the columns represent users. For any $k \in [K]$, if $p_{j,k}=\star$, then it implies that the subfiles $W^n_j,\forall n \in [N]$ are placed in the $k^{th}$ user's cache. If $p_{j,k}=s$ is an integer, it means that the user $k$ does not have access to the $j^{th}$ packet of any of the files. Condition $C1$ ensures that all users have access to some $Z$ subfiles of all the files. Thus, the memory constraint $M={NZ}/{F}$ is satisfied. In the delivery phase, for a demand vector $\mathbf{d}=(d_1,\ldots,d_K)$, the server sends messages of the form: 
\begin{equation}
  \underset{\substack{p_{j,k}=s \\j\in [F],  \textrm{\hspace{0.05cm}} k\in[K]}}{\bigoplus}W^{d_k}_j \textrm{\hspace{0.2cm}}, \forall s \in [S].
  \label{del_pda}
\end{equation}

Condition $C2$ and \eqref{del_pda} together imply that the number of messages transmitted by the server is $S$, and the delivery load required  is $\frac{S}{F}$. The decodability is guaranteed by condition $C3$.

\subsubsection{Generalized PDAs}

Generalized PDA was first introduced in \cite{PeR} to describe coded caching schemes for shared caches. It is defined as follows:
\begin{defn}
	(\cite{PeR})	For positive integers $K,F,Z,S$ and $I$, an $F \times K$ array $\mathbf{G}=(g_{j,k})$, $j \in [F]$, $k \in [K]$ composed of $\star $ and numerical entries from a subset of $[S] \times [I]$ where $[S] \times [I] :=\{(s,i) : s \in [S], i\in [I]\}$ is called a $(K,F,Z,[S] \times [I])$ generalized PDA if it satisfies the following conditions:
	\begin{enumerate}[label= C\arabic*.]
		\item The symbol $\star$ appears $Z$ times in each column.
		\item Each integer from the sets $\{1,2,\ldots,S\}$ and $\{1, 2,\ldots,I\}$ occurs at least once in the array.
		\item For any two distinct entries $g_{j_1,k_1}$ and $g_{j_2,k_2}$, $g_{j_1,k_1}=g_{j_2,k_2}=(s,i)$ is a numerical entry only if
		\begin{enumerate}[label=\alph*)]
			
			\item $j_1 \neq j_2$, $k_1 \neq k_2$, i.e., they lie in distinct rows and distinct columns, and
			\item $g_{j_1,k_2}=g_{j_2,k_1}=\star$, i.e., the corresponding $2\times2$ sub-array formed by rows $j_1, j_2$ and columns $k_1,k_2$ must be of the following form:
			\begin{center}
				$\begin{pmatrix}
				(s,i) & \star\\
				\star & (s,i)
				\end{pmatrix}$
				\hspace{0.3cm}or\hspace{0.3cm}
				$\begin{pmatrix}
				\star & (s,i) \\
				(s,i) & \star
				\end{pmatrix}$ 
				
			\end{center}
		\end{enumerate}
		\item For any four distinct entries $g_{j_1,k_1}$, $g_{j_1,k_2}$, $g_{j_2,k_1}$ and $g_{j_2,k_2}$, if $g_{j_1,k_1} =(s,i_1)$, $g_{j_2,k_1}=\star$ and  $g_{j_1,k_2}=(s,i_2)$, then $g_{j_2,k_2}=\star$. It is represented as follows:
		\begin{center}
			$\begin{pmatrix}
			(s,i_1) & (s,i_2)\\
			\star & \star
			\end{pmatrix}$
			
		\end{center}
		
	\end{enumerate}
	
\end{defn}

As the name suggests, generalized PDAs were a modified version of PDAs to accommodate the shared cache setting. We now describe how a generalized PDA represents a coded caching scheme for shared caches.

\begin{algorithm}[t]
	\renewcommand{\thealgorithm}{1}
	\caption{Construction of Generalized PDA for a given shared caching problem \cite{PeR}.}
	\label{alg:gpda_construct}
	\hspace*{\algorithmicindent} \textbf{Input}: $(\Lambda, F, Z, S)$ PDA $\mathbf{P}$, Number of users $K$,\\
	\hspace*{\algorithmicindent} \hspace{0.8cm} Association profile $\mathcal{L}=(\mathcal{L}_1,\mathcal{L}_2,\ldots,\mathcal{L}_{\Lambda})$.\\
	\hspace*{\algorithmicindent} \textbf{Output}: Generalized PDA $\mathbf{G} = (g_{j,k})_{F \times K}$
	\begin{algorithmic}[1]
		\State $k \gets 1$ 
		\For {$\lambda \in [\Lambda]$}
		\If {$\mathcal{L}_{\lambda}>0$}
		\For {$i \in [0,\mathcal{L}_{\lambda})$}
		\State $\mathbf{g}_k = \mathbf{p}_{\lambda}$
		\For {$j \in [F]$}
		\If {${g}_{j,k} \neq \star$ } 
		\State $g_{j,k}=(g_{j,k},1)$
		\State ${g}_{j,k} = {g}_{j,k}+(0,i)$.
		\EndIf
		\EndFor
		\State $k \gets k+1$		   		
		\EndFor
		\EndIf
		\EndFor
	\end{algorithmic}
\end{algorithm}

Consider a shared caching problem with $K$ users, $\Lambda$ caches, each of normalized size $\frac{M}{N}$. To obtain a scheme with less sub-packetization level for the above problem, we start with a $(\Lambda, F, Z, S)$ PDA that conform to $\frac{Z}{F}=\frac{M}{N}$. Each column in the PDA corresponds to a helper cache and each row, $j \in [F]$ represents subfiles $W^n_j, \forall n \in [N]$. The content placement is performed according to the `$\star$'s in the corresponding column. In \cite{PeR}, the user-to-cache association or association profile, $\mathcal{L}$ is known only after the placement phase. Once it is known, a $(K,F,Z,[S]\times[\mathcal{L}_1])$ generalized PDA $\mathbf{G}=(g_{j,k})$ is constructed using Algorithm~\ref{alg:gpda_construct} \cite{PeR}. Condition $C4$ ensures that the users connected to the same cache have the same side-information. For a demand vector $\mathbf{d}$, the delivery scheme is as follows: $$\underset{{g_{j,k}=(s,i)}}{\bigoplus}W^{d_k}_j, \textrm{\hspace{0.25cm}} \forall \textrm{\hspace{0.15cm}} (s,i) \in \mathbf{G}$$ where, $j \in [F]$ and $k \in [K]$. Thus, the delivery load required is obtained as:
\begin{equation}
	R(\mathcal{L}) = 	\frac{\displaystyle\sum_{s=1}^{S} \max\{i : (s,i) \textrm{ appears in } \mathbf{G}, i \in [\mathcal{L}_1] \}}{F}.
	\label{rate_exp}
\end{equation}
Or, equivalently
\begin{align}
R(\mathcal{L})  = 	\frac{\displaystyle\sum_{s=1}^{S} \mathcal{L}_{\tau_s} }{F}, \textrm{ \hspace{0.1cm}} \tau_s \triangleq \min\{\lambda \in [\Lambda], s \in \mathbf{p}_{\lambda} \}, \textrm{\hspace{0.1cm}} \forall s \in [S].
\label{rate_exp2}
\end{align}

The expression in \eqref{rate_exp2} follows from the fact that the maximum value of `$i$'  associated with each $s \in [S]$ depends on the column index, $\tau_s$ which corresponds to the most populated cache (cache serving maximum number of users) amongst those columns in which $s$ occurs in $\mathbf{P}$. Since the association profile, $\mathcal{L}$ is sorted in non-increasing order, $\tau_s$ is defined as 
$\min\{\lambda \in [\Lambda], s \in \mathbf{p}_{\lambda} \}$.

\begin{rem}
If each cache has got only one user accessing it, then the entries $(s,1)$ in the  $(K, F, Z, [S] \times [1])$ generalized PDA $\mathbf{G}$ can be replaced by $s$, and thus $\mathbf{G}$ reduces to a $(K, F, Z, S)$ PDA.
\end{rem}

\section{PDA based schemes for Shared Caches with User-to-Cache Association Profile known}
In this section, we first illustrate how the permutation of columns of the PDA that we begin with affect the delivery load, $R(\mathcal{L})$ required for a given shared caching problem. Then, we propose a set of rules to identify the PDA from the set of its equivalent PDAs which gives the best performance for the given problem.
\subsection{Motivating Example} 
\label{sec:motexmp}
Consider a shared cache network with $K=17$ users, $N=17$ files, $\Lambda=6$ caches, each with normalized size $\frac{M}{N}=\frac{1}{3}$. For this network, we choose a $(6,3,1,6)$ PDA $\mathbf{P}$ given in \eqref{pda}, such that 
$\frac{Z}{F}=\frac{M}{N}$ is satisfied. 

\begin{equation}
	\mathbf{P}={\begin{pmatrix}
			\star & 3 & 5 & \star & 1 & 2 \\
			1 & \star & 6 & 3 & \star & 4 \\
			2 & 4 & \star & 5 & 6 & \star
		\end{pmatrix}}
	\label{pda}
\end{equation}

Each column in $\mathbf{P}$ corresponds to a helper cache. The contents placed in cache $\lambda$ is denoted by $\mathcal{Z}_{\lambda}$. Then,
\begin{align*}
	\mathcal{Z}_1 & = \{W^n_{1}, \forall n \in [17]]\}, \mathcal{Z}_2 = \{W^n_{2}, \forall n \in [17]\} ,\\
	\mathcal{Z}_3 & = \{W^n_{3}, \forall n \in [17]\}, \mathcal{Z}_4 = \{W^n_{1}, \forall n \in [17]\} , \\
	\mathcal{Z}_5 & = \{W^n_{2}, \forall n \in [17]\} , \mathcal{Z}_6 = \{W^n_{3}, \forall n \in [17]\}.
\end{align*}

Let the user-to-cache association be such that $\mathcal{L}=(5,4,3,2,2,1)$. Then, construct a $(17, 3,1, [6] \times [5])$ generalized PDA $\mathbf{G}$ as described in Algorithm~\ref{alg:gpda_construct}. The array $\mathbf{G}$ is given in \eqref{gpda}. Each column in $\mathbf{G}$ corresponds to a user $k \in [17]$, and the `$\star$'s in the column represent the subfiles that each user has access to.

In the delivery phase, the server transmits a message corresponding to every distinct ordered pair $(s,i)$. Assume that the $k^{th}$ user demands the $k^{th}$ file, then the transmissions are as follows:
\begin{align*}
	X_{(1,1)} & = W^1_2 \oplus W^{15}_1, \textrm{\hspace{0.1cm}} X_{(1,2)} = W^2_2 \oplus W^{16}_1, \textrm{\hspace{0.1cm}} X_{(1,3)} = W^3_2 \\
	X_{(1,4)} & = W^4_2, \textrm{\hspace{0.1cm}} X_{(1,5)} = W^5_2, \textrm{\hspace{0.1cm}}  X_{(2,1)} = W^1_3 \oplus W^{17}_1, \textrm{\hspace{0.1cm}} \\
	X_{(2,2)} & = W^2_3, \textrm{\hspace{0.1cm}} X_{(2,3)} = W^3_3, \textrm{\hspace{0.1cm}} X_{(2,4)} = W^4_3, \textrm{\hspace{0.1cm}} X_{(2,5)} = W^5_3, \\
	X_{(3,1)} & = W^6_1 \oplus W^{13}_2, \textrm{\hspace{0.1cm}} X_{(3,2)} = W^7_1 \oplus W^{14}_2, \textrm{\hspace{0.1cm}} X_{(3,3)} = W^8_1, \textrm{\hspace{0.1cm}}\\
	X_{(3,4)} & = W^9_1, \textrm{\hspace{0.1cm}} X_{(4,1)} = W^6_3 \oplus W^{17}_2, \textrm{\hspace{0.1cm}} X_{(4,2)} = W^7_3, \textrm{\hspace{0.1cm}}\\
	X_{(4,3)} & = W^8_3, \textrm{\hspace{0.1cm}} X_{(4,4)}  = W^9_3, \textrm{\hspace{0.1cm}} X_{(5,1)}  = W^{10}_1 \oplus W^{13}_3, \\
	X_{(5,2)} &= W^{11}_1 \oplus W^{14}_3, \textrm{\hspace{0.1cm}} X_{(5,3)} = W^{12}_1, \\ X_{(6,1)} & = W^{10}_2 \oplus W^{15}_3, \textrm{\hspace{0.1cm}} X_{(6,2)} = W^{11}_2 \oplus W^{16}_3, \textrm{\hspace{0.1cm}}\\
	 X_{(6,3)} & = W^{12}_2.
\end{align*}

Thus, the sub-packetization level required is $3$ and the worst-case delivery load is 
\begin{align}
   R(\mathcal{L}=(5,4,3,2,2,1)) & = \frac{2(\mathcal{L}_1 + \mathcal{L}_2 + \mathcal{L}_3 )}{F} \label{rate_pda}\\ & = 24/3=8 \notag.
   \end{align}

Instead of $\mathbf{P}$, assume that we have started with $\mathbf{P}^{\prime}$ in \eqref{pda2}, which is also a $(6,3,1,6)$ PDA and, is related to $\mathbf{P}$ by column permutations. Hence, the arrays $\mathbf{P}$ and $\mathbf{P}^{\prime}$ are equivalent PDAs.

\begin{equation}
	\mathbf{P}^{\prime}={\begin{pmatrix}
			\star & 1 & 2 & 3 & 5 & \star \\
			1 & \star & 4 & \star & 6 & 3 \\
			2 & 6 & \star & 4 & \star & 5
		\end{pmatrix}}
	\label{pda2}
\end{equation}

For the same user-to-cache association and $\mathcal{L}$ considered in the previous case, we get the $(17,3,1,[6] \times [5])$ generalized PDA $\mathbf{G}^{\prime}$ as given in \eqref{gpda2}. The sub-packetization level required remains the same but the delivery load is given as

\begin{align}
R(\mathcal{L}=(5,4,3,2,2,1)) & = \frac{2\mathcal{L}_1 + \mathcal{L}_2  + \mathcal{L}_3 + \mathcal{L}_4 + \mathcal{L}_5}{F} \label{rate_pda2}\\ & = 21/3=7 \notag.
\end{align}

From the expressions in \eqref{rate_pda} and \eqref{rate_pda2}, it is straightforward that $R(\mathcal{L})$ in \eqref{rate_pda2} is smaller than the delivery load $R(\mathcal{L})$ in \eqref{rate_pda}. This observation is in stark contrast to the case with the coded caching schemes given by PDAs for dedicated cache networks. In a dedicated cache network model, the equivalent PDAs result in the same performance but change the content placement. Hence, for a given shared caching problem, our objective is to find the best possible arrangement of the columns of the PDA that is used. From the above discussed example, it follows that the association profile, $\mathcal{L}$ needs to be considered while placing the contents in the helper caches.

\setcounter{MaxMatrixCols}{20}
\begin{figure*}
	\begin{align}
		\mathbf{G} = \footnotesize{\begin{pmatrix}
				\star & \star & \star & \star & \star & (3,1) & (3,2) & (3,3) & (3,4) & (5,1) & (5,2) & (5,3) & \star & \star & (1,1) & (1,2) & (2,1) \\
				(1,1) & (1,2) & (1,3) & (1,4) & (1,5) & \star & \star & \star & \star & (6,1) & (6,2) & (6,3) & (3,1) & (3,2) & \star & \star & (4,1) \\
				(2,1) & (2,2) & (2,3) & (2,4) & (2,5) & (4,1) & (4,2) & (4,3) & (4,4) & \star & \star & \star & (5,1) & (5,2) & (6,1) & (6,2) & \star
		\end{pmatrix}}
		\label{gpda}
	\end{align}
\end{figure*}
\begin{figure*}
	\begin{align}
		\mathbf{G}^{\prime} = \footnotesize{\begin{pmatrix}
				\star & \star & \star & \star & \star & (1,1) & (1,2) & (1,3) & (1,4) & (2,1) & (2,2) & (2,3) & (3,1) & (3,2) & (5,1) & (5,2) & \star \\
				(1,1) & (1,2) & (1,3) & (1,4) & (1,5) & \star & \star & \star & \star & (4,1) & (4,2) & (4,3) & \star & \star & (6,1) & (6,2) & (3,1) \\
				(2,1) & (2,2) & (2,3) & (2,4) & (2,5) & (6,1) & (6,2) & (6,3) & (6,4) & \star & \star & \star & (4,1) & (4,2) & \star & \star & (5,1)
		\end{pmatrix}}
		\label{gpda2}
	\end{align}
\end{figure*}

\subsection{Procedure to identify the PDA with the best performance }
\label{subsec:proc}

Consider a shared caching problem with $K$ users, $\Lambda$ caches each of normalized size $M/N$ and association profile, $\mathcal{L}$ known a priori. To obtain a scheme for the above, we have taken a $(\Lambda, F, Z, S)$ PDA $\mathbf{P}$ with $\frac{Z}{F}=\frac{M}{N}$. The value of `$F$' is within the allowable sub-packetization level. If the association profile $\mathcal{L}$ is not in the sorted order, relabel and rearrange the columns of $\mathbf{P}$ such that the caches are in the non-increasing order of its occupancy. With slight abuse of notation, we continue to call the rearranged array as $\mathbf{P}$ itself. The delivery load expression in \eqref{rate_pda2} gives some insights on how to find the best arrangement of the columns of $\mathbf{P}$. The expression in \eqref{rate_pda2} indicates that if some of the integers $s \in [S]$ make its first occurrence in the columns corresponding to the caches connected with less number of users, then the number of transmissions gets reduced. This is well illustrated by the example given in Section~\ref{sec:motexmp}. Inspired by this observation, we give a general rule of thumb to arrive at the best column permutation of $\mathbf{P}$ in the sequel. 

Define $\mathcal{I}_{\lambda}$ as the set of distinct integers appeared till $\lambda^{th}$ column in $\mathbf{P}$, where $\lambda \in [\Lambda]$. Obviously, $|\mathcal{I}_1| = (F-Z)$ and $|\mathcal{I}_{\Lambda}|=S$. In a $(\Lambda,F,Z,S)$ PDA, it is possible that all the $S$ integers could have occurred before the last column. Let $\alpha$ be the first column index such that $|\mathcal{I}_{\alpha}|=S$, then, $|\mathcal{I}_{\alpha +1}| = |\mathcal{I}_{\alpha +2}| = \ldots = |\mathcal{I}_{\Lambda}| = S$. We need `$\alpha$' to be maximized. Consequently, the number of new integers appearing in the columns corresponding to more populated caches gets decreased, thereby gaining in the number of transmissions needed. Denote $\alpha^{*}$ as the first column index by which all the integers have appeared in the optimal arrangement. Then, $\alpha < \alpha^{*}$, for any other arrangement of columns. 

The parameter $\tau_s$ in \eqref{rate_pda2} can be viewed as the column index in which the integer $s$ first appears in $\mathbf{P}$, if $\mathcal{L}$ is in the sorted order. If an integer $s \in [S]$ occurs $g_s$ times in $\mathbf{P}$, then $\tau_s \leq K-g_s+1 $. While maximizing $\alpha$, effectively, we are maximizing certain $\tau_s$ values. The PDA obtained after rearranging the columns in an optimal way is denoted by $\widehat{\mathbf{P}}$. The procedure to find $\widehat{\mathbf{P}}$ is given below.

\begin{enumerate}
	\item Choose a pair of columns $(\lambda_1, \lambda_2)$ from the PDA $\mathbf{P}$ and, find the intersection between the set of integers present in the columns, $\lambda_1$ and $\lambda_2$. Repeat this process for all the $\binom{\Lambda}{2}$ pairs. Let the cardinality of the intersection set be called as intersection number. Then, select the pair, $(\lambda_1,\lambda_2)$ which gives the largest intersection number and, assign it as the first two columns, $\widehat{\mathbf{p}_1}$ and $\widehat{\mathbf{p}_2}$.

	\item Find $\mathcal{I}_2$, which is the union of the set of integers present in columns $\widehat{\mathbf{p}_1}$ and $\widehat{\mathbf{p}_2}$. That is,
	$$ \mathcal{I}_2 = \underset{l \in \{1,2\}}{\bigcup} \{s: \widehat{p}_{j,l}=s, j \in [F]\}. $$
	
	If $|\mathcal{I}_2|=S$, arrange the remaining columns in any random order to obtain $\widehat{\mathbf{P}}$. Else, pick each column from the remaining $\Lambda-2$ columns of $\mathbf{P}$ and find its intersection with $\mathcal{I}_2$. The column that gives the largest intersection number is assigned as $\widehat{\mathbf{p}_3}$.

    \item To identify the remaining columns, $\widehat{\mathbf{p}_l}$ where $l \in \{4,5,\ldots,\Lambda\}$, first obtain
    $$\mathcal{I}_{l-1} = \mathcal{I}_{l-2} \textrm{\hspace{0.1cm}}\bigcup \textrm{\hspace{0.1cm}} \{s: \widehat{p}_{j,l-1}=s, j \in [F]\}.$$ 
    
    Check if $|\mathcal{I}_{l-1}|=S$. If not, choose each column from the remaining $\Lambda-l+1$ columns and find its intersection with $\mathcal{I}_{l-1}$. The column with the largest intersection number is taken as $\widehat{\mathbf{p}_l}$. This step is repeated till all $s \in [S]$ appears in $\widehat{\mathbf{P}}$. Once all the integers have appeared in $\widehat{\mathbf{P}}$, the rest of the columns of $\mathbf{P}$ can be put arbitrarily as $R(\mathcal{L})$ depends only on $\tau_s$, $\forall s \in [S] $.
    
 \end{enumerate}

In the end, we obtain $\widehat{\mathbf{P}}$, which is equivalent to $\mathbf{P}$. In any arrangement or ordering, $|\mathcal{I}_1|=F-Z$. Hence, there is always $\mathcal{L}_1$ transmissions associated with every integer present in the first column. As mentioned before, to gain in $R(\mathcal{L})$, the number of new integers occurring in $\widehat{\mathbf{p}_2}$ needs to be reduced, and this is the reason behind designing Step $1$ in the given way. In Steps $2$ and $3$ also, we do the same by maximizing the intersection number. To summarize, the three steps mentioned above ensure that in $\widehat{\mathbf{P}}$, new integers appear more towards the columns corresponding to less occupied caches. The intuition and significance behind it is that the coding gain, defined as the number of users benefiting from a transmission, increases compared to the delivery policy obtained using $\mathbf{P}$ (evident from the transmissions given by $\mathbf{G}$ and $\mathbf{G}^{\prime}$ in the example in Section~\ref{sec:motexmp}).

 Note that the array $\widehat{\mathbf{P}}$ obtained at the end of the above procedure is optimal only if the column that results in the largest intersection number is unique. If there is more than one column with the largest intersection number, then arbitrarily picking one does not work in general. In that case, the decision depends on the intersection numbers that are going to obtain at the successive stages. Finding the subsequent intersection numbers for each available option, and then making a decision becomes computationally tedious as $\Lambda$ scales. But for certain specific PDA constructions, we could identify a rule on how to get the best arrangement. For instance, Construction B given in \cite{YCT}. 
 
 \subsection{Finding the optimal ordering for PDAs obtained using Construction B in \cite{YCT}}
 \label{subsec:proc_constB}
 
 First, we briefly review Construction B. It gives a $(q(m+1),(q-1)q^m,(q-1)q^{m-2},q^m)$ PDA $\mathbf{P}$ where $q, m \in \mathbb{N}$ and $q \geq 2$. In $\mathbf{P}$, each integer occurs $(q-1)(m+1)$ times and $\frac{Z}{F}=\frac{q-1}{q}$.  The rows of $\mathbf{P}$ are indexed using $(m+1)$ tuples $(j_m,\ldots,j_1,j_0)_q$ over $\mathbb{F}_q$, where $j_m \in [0,q-1)$ and $j_0,j_1,\ldots,j_{m-1} \in [0,q)$. The columns are indexed using an ordered pair $(u,v)$ where, $u \in [0,m]$ and $v \in [0,q)$. The $q(m+1)$ columns can be split into $m+1$ sets with each set containing $q$ columns and each set is being identified by the value of $u$. 
 
 The symbol `$\star$' is placed using the concept of partitions. Our main focus is on the integers of $\mathbf{P}$, which are represented using $m$ tuples $(s_{m-1},\ldots,s_1,s_0)_q$ over $\mathbb{F}_q$. Thus, we have $q^m$ integers in total.
 In columns $(u,v)$ where $u \in [0,m)$ and $v \in [0,q)$, all $m$ tuples $(s_{m-1},\ldots,s_1,s_0)_q$ are present except those with $s_u =v$. In columns $(u=m,v)$, all those tuples satisfying the condition $(\sum_{i=0}^{m-1}s_i)_q = v-1$ are present. 
 
 The PDAs obtained from Construction B \cite{YCT} is suitable for shared cache systems with $\Lambda$ expressible as $q(m+1)$ and $\frac{M}{N}=\frac{q-1}{q}$, where $q \geq 2$. Consider such a shared caching problem with $K$ users and association profile, $\mathcal{L}=(\mathcal{L}_1,\mathcal{L}_2,\ldots,\mathcal{L}_{\Lambda})$ known. Assume that $\mathcal{L}$ is already present in the sorted order.
 
As mentioned earlier, split the $q(m+1)$ columns into $(m+1)$ sets based on the value of $u$. First, pick a column from any one of these $(m+1)$ sets. Let $(u_1,v_1)$ be the index of the chosen column. The second column should be picked from any remaining $m$ sets because we aim to find the column with the largest intersection number. Therefore, it should not be from the set with $u=u_1$, which follows from the PDA construction. Choose any one column arbitrarily from the $mq$ options available. The successive $(m-2)$ columns should also be picked from different sets and, this is possible as there are $m+1$ sets. By doing so, the intersection number gets maximized. Since there are many choices available, randomly pick one as done in the previous case and, it does not affect the performance. Let the indices of the columns chosen so far be $(u_1,v_1)$, $(u_2,v_2),\ldots,(u_m,v_m)$. Then, $u_1 \neq u_2 \neq \ldots \neq u_m$. The set of integers or $m$ tuples appeared so far are given by $\mathcal{I}_m$. For ease of exposition, assume that $u_1=0, u_2=1, \ldots, u_m =m-1$. Then, 
$$\mathcal{I}_m = [0,q^m) \backslash (v_m,\ldots,v_2,v_1)_q. $$
  The set $\mathcal{I}_m$ contains all the $m$ tuples except $(v_m,\ldots,v_2,v_1)_q$, hence, $|\mathcal{I}_m|=q^m-1$. As done previously, the $(m+1)^{th}$ column needs to be chosen from the remaining one set. In that set, among the $q$ columns, there is only one column which does not have the vector $(v_m,\ldots,v_2,v_1)$. Hence, pick that column as the $(m+1)^{th}$ column. Then, $\mathcal{I}_m = \mathcal{I}_{m+1}$. The remaining columns can be picked in any way because all those columns contain $(v_m,\ldots,v_2,v_1)$. Thus, $\alpha^{*}=m+2$. Effectively, we are proceeding in the same way as described in the three steps given earlier in this section. When we get more than one option at any stage, arbitrarily picking one column helps in Construction B. But it is not true in all the other constructions, for example, Construction A given in \cite{YCT}. 
  
  \begin{rem}
  	In a $g$-regular PDA, all the integers $s \in [S]$ occur $g$ times in the PDA. Then, $\tau_s \leq K-g+1$, $\forall s \in [S]$. In the optimal arrangement, there exists at least one $s \in [S]$ such that $\tau_s = K-g+1$. Then, it implies $\alpha^{*}=(K-g+1)$. For example, in Construction B, $g=(q-1)(m+1)$. In the optimal ordering, for $s=(v_m,\ldots,v_2,v_1)_q$, we obtained $\tau_{s}= K-g+1 = m+2$.
  \end{rem}
  After ordering the columns of $\mathbf{P}$, we get a PDA $\widehat{\mathbf{P}}$ equivalent to $\mathbf{P}$. The columns of $\widehat{\mathbf{P}}$ represent the helper caches arranged in the non-increasing order of their occupancy number. Then, proceed with $\widehat{\mathbf{P}}$ to obtain the corresponding shared cache scheme.

 \textit{Performance measures}: The sub-packetization level required in the obtained scheme is $(q-1)q^m$. Using the expression in \eqref{rate_exp2}, the delivery load $R(\mathcal{L})$ obtained with $\widehat{\mathbf{P}}$ is:
 \begin{equation}
  R(\mathcal{L}) =  \frac{\mathcal{L}_1}{q}+ \frac{\mathcal{L}_2}{q^2}+ \ldots+\frac{\mathcal{L}_m}{q^m}+ \frac{\mathcal{L}_{m+2}}{(q-1)q^m}.
 \label{rate_constB}
 \end{equation}
 
 The expression in \eqref{rate_constB} is derived as follows: the first term in the summand comes from the $F-Z=(q-1)q^{m-1}$ integers in the first column. Since the first $m+1$ columns in $\widehat{\mathbf{P}}$ are picked from $m+1$ different sets, the number of new integers appearing in each column $\widehat{\mathbf{p}_i}$, $i \in [m]$ can be written as $(q-1)q^{m-i}$. This leads to the first $m$ terms in the $R(\mathcal{L})$ expression in \eqref{rate_constB}. Since there is no new integer appearing in the $(m+1)^{th}$ column, $|\mathcal{I}_m|=|\mathcal{I}_{m+1}|=q^m-1$. The last term in \eqref{rate_constB} corresponds to the remaining one integer $s$ with $\tau_s = m+2$. The $R(\mathcal{L})$ in \eqref{rate_constB} is the best or the minimum delivery load that we could achieve for a shared caching scheme using a PDA derived from Construction B.

 If we have employed the PDA $\mathbf{P}$ directly without ordering, then $R(\mathcal{L})$ is obtained as:
 
\begin{equation}
  R(\mathcal{L}) = \frac{\mathcal{L}_1}{q} + \frac{\mathcal{L}_2}{q(q-1)}.
  \label{rate_PDAworst}
\end{equation}
 
 The delivery load obtained in \eqref{rate_PDAworst} is greater than that in \eqref{rate_constB}. For association profiles with $\mathcal{L}_2=\mathcal{L}_3=\ldots=\mathcal{L}_{m+1}=\mathcal{L}_{m+2}$, the expressions in \eqref{rate_constB} and \eqref{rate_PDAworst} match. That is, we obtain the same performance with $\mathbf{P}$ and $\widehat{\mathbf{P}}$.

 \begin{table*}[ht]
 	\centering
 	\caption{$(9,18,12,9)$ PDA $\mathbf{P}$ obtained using Construction B ($q=3$, $m=2$) \cite{YCT}.}
 	\label{tab:pda}
 	\begin{tabular}[t]{|c|| c c c | c c c | c c c|}
 		
 		\hline 
 		
 		$(j_2,j_0,j_1)_3 \backslash (u,v)$ & $(0,0)$ & $(0,1)$ &$(0,2)$ & $(1,0)$ & $(1,1)$ & $(1,2)$ & $(2,0)$ & $(2,1)$ & $(2,2)$  \\
 		\hline \hline 	\rule{0pt}{3ex}
 		
 		$(0,0,0)_3$ & $(0,1)_3$ & $\star$ & $\star$ & $(1,0)_3$ & $\star$ & $\star$ & $\star$ & $(0,0)_3$ & $\star$ \\ 
 		$(0,0,1)_3$ & $\star$ & $(0,2)_3$ & $\star$ & $(1,1)_3$ & $\star$ & $\star$ & $\star$ & $\star$ & $(0,1)_3$ \\
 		$(0,0,2)_3$ & $\star$ & $\star$ & $(0,0)_3$ & $(1,2)_3$ & $\star$ & $\star$ & $(0,2)_3$ & $\star$ & $\star$ \\
 		$(0,1,0)_3$ & $(1,1)_3$ & $\star$ & $\star$ & $\star$ & $(2,0)_3$ & $\star$ & $\star$ & $\star$ & $(1,0)_3$ \\
 		$(0,1,1)_3$ & $\star$ & $(1,2)_3$ & $\star$ & $\star$ & $(2,1)_3$ & $\star$ & $(1,1)_3$ & $\star$ & $\star$ \\
 		$(0,1,2)_3$ & $\star$ & $\star$ & $(1,0)_3$ & $\star$ & $(2,2)_3$ & $\star$ & $\star$ & $(1,2)_3$ & $\star$ \\
 		$(0,2,0)_3$ & $(2,1)_3$ & $\star$ & $\star$ & $\star$ & $\star$ & $(0,0)_3$ & $(2,0)_3$ & $\star$ & $\star$ \\
 		$(0,2,1)_3$ & $\star$ & $(2,2)_3$ & $\star$ & $\star$ & $\star$ & $(0,1)_3$ & $\star$ & $(2,1)_3$ & $\star$ \\
 		$(0,2,2)_3$ & $\star$ & $\star$ & $(2,0)_3$ & $\star$ & $\star$ & $(0,2)_3$ & $\star$ & $\star$ & $(2,2)_3$ \\
 		$(1,0,0)_3$ & $(0,2)_3$ & $\star$ & $\star$ & $(2,0)_3$ & $\star$ & $\star$ & $\star$ & $\star$ & $(0,0)_3$ \\
 		$(1,0,1)_3$ & $\star$ & $(0,0)_3$ & $\star$ & $(2,1)_3$ & $\star$ & $\star$ & $(0,1)_3$ & $\star$ & $\star$ \\
 		$(1,0,2)_3$ & $\star$ & $\star$ & $(0,1)_3$ & $(2,2)_3$ & $\star$ & $\star$ & $\star$ & $(0,2)_3$ & $\star$ \\
 		$(1,1,0)_3$ & $(1,2)_3$ & $\star$ & $\star$ & $\star$ & $(0,0)_3$ & $\star$ & $(1,0)_3$ & $\star$ & $\star$ \\
 		$(1,1,1)_3$ & $\star$ & $(1,0)_3$ & $\star$ & $\star$ & $(0,1)_3$ & $\star$ & $\star$ & $(1,1)_3$ & $\star$ \\
 		$(1,1,2)_3$ & $\star$ & $\star$ & $(1,1)_3$ & $\star$ & $(0,2)_3$ & $\star$ & $\star$ & $\star$ & $(1,2)_3$ \\
 		$(1,2,0)_3$ & $(2,2)_3$ & $\star$ & $\star$ & $\star$ & $\star$ & $(1,0)_3$ & $\star$ & $(2,0)_3$ & $\star$ \\
 		$(1,2,1)_3$ & $\star$ & $(2,0)_3$ & $\star$ & $\star$ & $\star$ & $(1,1)_3$ & $\star$ & $\star$ & $(2,1)_3$ \\
 		$(1,2,2)_3$ & $\star$ & $\star$ & $(2,1)_3$ & $\star$ & $\star$ & $(1,2)_3$ & $(2,2)_3$ & $\star$ & $\star$ \\
 		\hline
 	\end{tabular}
 	
 \end{table*}

 \begin{table*}[h]
	\centering
	\caption{$(9,18,12,9)$ PDA $\widehat{\mathbf{P}}$ which is equivalent to $\mathbf{P}$.}
	\label{tab:pdaeq}
	\begin{tabular}[t]{|c|| c c c  c c c  c c c|}
		
		\hline 
		
		$(j_2,j_0,j_1)_3 \backslash (u,v)$ & $(0,0)$ & $(1,0)$ &$(2,0)$ & $(0,1)$ & $(0,2)$ & $(1,1)$ & $(1,2)$ & $(2,1)$ & $(2,2)$  \\
		\hline \hline 	\rule{0pt}{3ex}
		
		$(0,0,0)_3$ & $(0,1)_3$  & $(1,0)_3$ & $\star$ & $\star$ & $\star$ & $\star$ & $\star$ &  $(0,0)_3$ & $\star$ \\ 
		$(0,0,1)_3$ & $\star$  & $(1,1)_3$  & $\star$  & $(0,2)_3$ & $\star$ & $\star$ & $\star$ & $\star$ & $(0,1)_3$ \\
		$(0,0,2)_3$ & $\star$  & $(1,2)_3$  & $(0,2)_3$  & $\star$ & $(0,0)_3$ &  $\star$ & $\star$ & $\star$ & $\star$ \\
		$(0,1,0)_3$ & $(1,1)_3$ & $\star$ & $\star$  & $\star$ & $\star$ &  $(2,0)_3$ & $\star$ & $\star$ & $(1,0)_3$ \\
		$(0,1,1)_3$ & $\star$  & $\star$  & $(1,1)_3$ & $(1,2)_3$ & $\star$ &  $(2,1)_3$ & $\star$ &  $\star$ & $\star$ \\
		$(0,1,2)_3$ & $\star$ & $\star$  & $\star$  & $\star$ & $(1,0)_3$ &  $(2,2)_3$ & $\star$ & $(1,2)_3$ & $\star$ \\
		$(0,2,0)_3$ & $(2,1)_3$ & $\star$  & $(2,0)_3$ & $\star$ & $\star$ & $\star$ & $(0,0)_3$ &  $\star$ & $\star$ \\
		$(0,2,1)_3$ & $\star$  & $\star$  & $\star$ & $(2,2)_3$ & $\star$ & $\star$ & $(0,1)_3$ & $(2,1)_3$ & $\star$ \\
		$(0,2,2)_3$ & $\star$  & $\star$  & $\star$ & $\star$ & $(2,0)_3$ & $\star$ & $(0,2)_3$ & $\star$ & $(2,2)_3$ \\
		$(1,0,0)_3$ & $(0,2)_3$  & $(2,0)_3$ & $\star$ & $\star$ & $\star$ &  $\star$ & $\star$ & $\star$ & $(0,0)_3$ \\
		$(1,0,1)_3$ & $\star$  & $(2,1)_3$  & $(0,1)_3$ & $(0,0)_3$ & $\star$ &  $\star$ & $\star$ & $\star$ & $\star$ \\
		$(1,0,2)_3$ & $\star$ & $(2,2)_3$  & $\star$ & $\star$ & $(0,1)_3$ &  $\star$ & $\star$ & $(0,2)_3$ & $\star$ \\
		$(1,1,0)_3$ & $(1,2)_3$  & $\star$ & $(1,0)_3$  & $\star$ & $\star$ & $(0,0)_3$ & $\star$ &  $\star$ & $\star$ \\
		$(1,1,1)_3$ & $\star$  & $\star$  & $\star$  & $(1,0)_3$ & $\star$ &  $(0,1)_3$ & $\star$ & $(1,1)_3$ & $\star$ \\
		$(1,1,2)_3$ & $\star$ & $\star$ & $\star$  & $\star$ & $(1,1)_3$ &  $(0,2)_3$ & $\star$ & $\star$ & $(1,2)_3$ \\
		$(1,2,0)_3$ & $(2,2)_3$  & $\star$ & $\star$ & $\star$ & $\star$ &  $\star$ & $(1,0)_3$ & $(2,0)_3$ & $\star$ \\
		$(1,2,1)_3$ & $\star$  & $\star$  & $\star$ & $(2,0)_3$ & $\star$ & $\star$ & $(1,1)_3$ & $\star$ & $(2,1)_3$ \\
		$(1,2,2)_3$ & $\star$ & $\star$ & $(2,2)_3$ & $\star$ & $(2,1)_3$ & $\star$ & $(1,2)_3$ & $\star$ & $\star$ \\
		\hline
	\end{tabular}
	
\end{table*}

 Next, we present an example that clearly illustrates all the above discussed procedure. 

\textit{Example 2:} Consider a shared caching problem with $\Lambda=9$ helper caches, each of normalized size ${M}/{N}=2/3$ and $K=110$ users. The association profile is $\mathcal{L}=(30,25,20,10,8,5,5,4,3)$. The $(9,18,12,9)$ PDA $\mathbf{P}$ (obtained using Construction B) given in Table~\ref{tab:pda} can be used for this problem. The corresponding values of $q$ and $m$ are $3$ and $2$, respectively. The columns of $\mathbf{P}$ are split into $m+1=3$ sets. Each set is being identified with the value of $u$ and is comprised of $q=3$ columns. Each  column, $(u,v)$ from the sets $u \in \{0,1\}$ has all the vectors $(s_1,s_0)_3$ such that $s_u \neq v$. When $u=m$, each vector $(s_1,s_0)_3$ in the column, $(u,v)$ satisfies $(s_1+s_0)_3 = v-1$.

To obtain $\widehat{\mathbf{P}}$, first pick a column from any of the three sets. We choose $\mathbf{p}_1$ which belongs to the set $u=0$ and is indexed as $(0,0)$. Therefore, $\widehat{\mathbf{p}_1}$ is $(0,0)$. Observe that the set $\mathcal{I}_1$ contains all the vectors of the form $(s_1,s_0)_3$ with $s_0 \neq 0$. If the second column is picked from the same set with $u=0$, we get the intersection number as $3$ and, all the nine integers get appeared by the second column, i.e, $|\mathcal{I}_2|=9$. Whereas, if we pick a column from any of the other two sets, the intersection number obtained is $4$ and $|\mathcal{I}_2|=8$. Therefore, pick one column arbitrarily from the sets with $u=1$ or $u=2$. We choose $\mathbf{p}_4$ which is indexed as $(1,0)$. Therefore, $\widehat{\mathbf{p}_2}$ is $(1,0)$.
Then, $\mathcal{I}_2$ contains all the vectors $(s_1,s_0)_3$ except $(0,0)_3$. The third column needs to be taken from the remaining one set, which corresponds to $u=3$. Since the objective is to maximize the intersection number, the column that does not contain $(0,0)_3$ needs to be picked. Hence, the column $\mathbf{p}_7$ indexed as $(2,0)$ is chosen. Therefore, $\widehat{\mathbf{p}_3}$ is $(2,0)$. The remaining six columns of $\mathbf{P}$ can be arranged in any order as it all contains the vector $(0,0)_3$. Thus, the columns of $\widehat{\mathbf{P}}$ are as follows: $\widehat{\mathbf{p}_1} = (0,0) $, $\widehat{\mathbf{p}_2} = (1,0) $, $\widehat{\mathbf{p}_3} = (2,0) $, $\widehat{\mathbf{p}_4} = (0,1) $, $\widehat{\mathbf{p}_5} = (0,2) $, $\widehat{\mathbf{p}_6} = (1,1) $, $\widehat{\mathbf{p}_7} = (1,2) $, $\widehat{\mathbf{p}_8} = (2,1) $, $\widehat{\mathbf{p}_9} = (2,2) $. The array $\widehat{\mathbf{P}}$ is given in Table~\ref{tab:pdaeq}. Once $\widehat{\mathbf{P}}$ is obtained, the ternary representation of the integers are changed to decimal format before proceeding further. For the given association profile, $\mathcal{L}=(30,25,20,10,8,5,5,4,3)$, the equivalent PDA $\widehat{\mathbf{P}}$ in Table~\ref{tab:pdaeq} gives the best performance. The construction of GPDA and the transmissions are not illustrated here for the sake of brevity. We directly look at the performance of the obtained scheme.

The sub-packetization level required in the obtained scheme is $18$. The delivery load required by using $\widehat{\mathbf{P}}$ is $R(\mathcal{L})=\frac{30.6 + 25.2 + 10.1}{18}= 240/18 \approx 13.333$. Instead, if we start with ${\mathbf{P}}$, $R(\mathcal{L})=\frac{30.6 + 25.3}{18}=275/18 \approx 15.278$. Thus, we could gain in $R(\mathcal{L})$ just by rearranging the columns of the PDA without incurring any penalty. For this network parameters, the optimal scheme in \cite{PUE} requires the sub-packetization level, $F=84$ and the delivery load, $R(\mathcal{L})\approx 12.321$. Thus, by rearranging the columns of the PDA, the performance of the PDA derived scheme moves closer to that of the optimal scheme.

 \subsection{Effect of rearranging the columns on Maddah-Ali Niesen PDA (MN PDA)}
The PDA that represents the Maddah-Ali Niesen scheme \cite{YCT} is often called as MN PDA. The MN PDA for a $(K,M,N)$ coded caching scheme is defined as $(K, \binom{K}{t}, \binom{K-1}{t-1}, \binom{K}{t+1})$, where $t \triangleq \frac{KM}{N}$. As discussed in \cite{PeR}, for a given shared caching problem, the scheme in \cite{PUE} can be recovered using an MN PDA defined for $\Lambda$ users, which is given as $(\Lambda, \binom{\Lambda}{t}, \binom{\Lambda-1}{t-1}, \binom{\Lambda}{t+1})$.

Consider a shared caching problem where we start with an MN PDA $\mathbf{Q}$. When the aforementioned $3$ steps are performed on $\mathbf{Q}$, we obtain more than one option for the column with the largest intersection number. We can randomly choose one column whenever we confront such situations with $\mathbf{Q}$. The performance obtained by using $\mathbf{Q}$ or any other PDA that is equivalent to $\mathbf{Q}$ is same, as the scheme in \cite{PUE} is optimal under uncoded placement. Hence, there is no need to order the columns of an MN PDA according to the association profile, $\mathcal{L}$ as it does not improve $R(\mathcal{L})$.

 \begin{rem}
 	 To find the best arrangement of the columns, the exact user-to-cache association profile is not required for content placement. Instead, we need to know only the order of the helper caches in terms of the number of users served by it. That is, our proposed procedure needs only the knowledge about which cache serves the highest number of users, the second-highest number of users, and so on (the exact number $\mathcal{L}_i$, $i \in [\Lambda]$ is not essential). It is possible to obtain this order from statistical knowledge in many practical scenarios. 
 	 
\end{rem}

\section{Discussion and Related Work}
 For comparison, we considered only the scheme in \cite{PeR} and the optimal scheme (under uncoded placement) in \cite{PUE}. There are other centralized shared cache schemes in the literature \cite{IZY}, \cite{PaE}. The scheme in \cite{IZY} caches both uncoded and coded portions of files by taking into account the association profile. The association profile, $\mathcal{L}$ should be known exactly at the placement phase itself and, the scheme outperforms the optimal scheme in \cite{PUE} significantly if $\mathcal{L}$ is highly skewed. But, the scheme in \cite{IZY} finds the parameters of the caching scheme (for example, size of the subfiles) by solving a linear program. In \cite{PaE}, contrary to the conventional shared cache setting, the authors assumed a cumulative memory constraint and assigned heterogeneous memory sizes for each helper cache depending on the number of users connected to it. By following an optimized memory allocation across the caches, the scheme proposed in \cite{PaE} outperforms the scheme in \cite{PUE} but the sub-packetization level required is larger than that is needed in \cite{PUE}.

\section{Conclusion}

In this work, we considered the coded caching schemes for shared caches obtained from PDAs. The PDA that we choose to start with determines the placement and delivery policies, as each column represents a helper cache. Therefore, instead of following an association profile independent placement, we make use of the knowledge about the occupancy number of each helper cache and, accordingly, order the columns of the PDA. The concept of equivalent PDAs is introduced, and a general rule of thumb is proposed on finding the best ordering. The main point to emphasize here is that the equivalent PDAs need not result in the same delivery load for a given association profile. Hence, it is imperative to order the columns of the PDA that we choose, before proceeding further.

\section*{Acknowledgement}
This work was supported partly by the Science and Engineering Research Board (SERB) of Department of Science and Technology (DST), Government of India, through J.C Bose National Fellowship to Prof. B. Sundar Rajan.

\end{document}